# Multi-Frequency Magnonic Logic Circuits for Parallel Data Processing

Alexander Khitun
Electrical Engineering Department,
University of California Riverside, California 92521

**Abstract**

We describe and analyze magnonic logic circuits enabling parallel data processing on multiple frequencies. The circuits combine bi-stable (digital) input/output elements and an analog core. The data transmission and processing within the analog part is accomplished by the spin waves, where logic 0 and 1 are encoded into the phase of the propagating wave. The latter makes it possible to utilize a number of bit carrying frequencies as independent information channels. The operation of the magnonic logic circuits is illustrated by numerical modeling. We also present the estimates on the potential functional throughput enhancement and compare it with scaled CMOS. The described multi-frequency approach offers a fundamental advantage over the transistor-based circuitry and may provide an extra dimension for the Moor's law continuation. The shortcoming and potentials issues are also discussed.

**Introduction**

Moderns logic circuits consist of a large number of transistors fabricated on the surface of silicon wafer and interconnected by metallic wires. The transistors are arranged to perform Boolean operations (e.g. NOT, AND logic gates). Within this approach, the computational power is measured in the number of operations per time per area. In the past sixty years, a strait forward approach to functional throughput enhancement was associated with the increase of the number of transistors, which is well known as the Moor's law [1]. Decades of transistor-based circuitry perfection resulted in the Complementary Metal–Oxide–Semiconductor (CMOS) technology, which is the basis for current semiconductor industry. Unfortunately, CMOS technology is close to the fundamental limits mainly due to the power dissipation problems [2]. The latter stimulates a big deal of interest to the post-CMOS technologies able to overcome the current constrains.

The most of the "beyond CMOS" proposals are focused on the development of a new switch – a more efficient transistor [3]. There is still some room for the semiconductor transistors improvement by implementing novel materials (e.g. nano-tubes [4] , graphene [5], tunneling-based transistors [6], etc.). However, it is difficult to expect that the introduction of a new material may extend the Moor's law for multiple generations as it used to work for CMOS. At this moment, it is important to identify possible routes to alternative (possibly transistor-less) logic devices which may lead to a more powerful logic circuitry and offer a pathway for a long-term development. There are several fundamental constrains inherent to the transistor based logic devices: (i) logic variable is a scalar value (voltage), (ii) metallic interconnects do not perform any functional work on data processing, (iii) the transistor-based approach is volatile, requiring a permanent power supply even no computation is performed, and (iv) one transistor can process only *one bit per time*.  Addressing these fundamental issues is the key leading to a more efficient logic circuitry.

Magnonic logic circuits exploiting magnetization as a state variable and spin waves for information transmission and processing is one of the possible solutions. The basic concept of the magnonic logic circuits have been described in the preceding works [7-10]. The utilization of waves for data transmission makes it possible to code logic 0 and 1 in the phase of the propagating wave, use waveguides as passive logic elements for phase modulation and exploit wave interference for achieving logic functionality.  In this work, we describe and analyze the possibility of multi-frequency operation, where each frequency can serve as an independent information channel for information transmission and processing. The rest of the paper is organized as follows. In the next Section, we describe the basic elements and the principle of operation of a single-frequency magnonic logic circuit. Next, we extend consideration to the multi-frequency logic units. The operation of magnonic logic circuits is illustrated by numerical modeling in Section III. The advantages and limits of the multi-frequencies approach are discussed in the Sections IV and V, respectively.

## II. Principle of Operation and Basic Elements

We start with a description of the single-frequency operating magnonic logic circuit schematically shown in Fig.1. The circuit comprises the following elements: (i) magneto-electric cells, (ii) magnetic waveguides –spin wave buses, and a (iii) phase shifter. Magneto-electric cell (ME cell) is the element aimed to convert voltage pulses into the spin wave as well as to read-out the voltage produced by the spin waves.  The operation of the ME cell is based on the effect of magneto-electric coupling enabling magnetization control by applying an electric field and vice versa. ME cell can be realized by utilizing multiferroic materials to be discussed latter in the text. The waveguides are simply the strips of ferromagnetic material (e.g. NiFe) aimed to transmit the spin wave signals. The phase shifter is a passive element providing a $\pi$-phase shift to the propagating spin waves.

The principle of operation is the following. Initial information is received in the form of voltage pulses.  Input 0 and 1 are encoded in the polarity of the voltage applied to the input ME cells (e.g. +10mV correspond to logic state 0, and -10mV correspond to logic 1). The polarity of the applied voltage defines the initial phase of the spin wave signal (e.g. positive voltage results in the clockwise magnetization rotation and negative voltage results in the counter clockwise magnetization rotation). Thus, the input information is translated into the phase of the excited wave (e.g. initial phase 0 corresponds to logic state 0, and initial phase $\pi$ corresponds to logic 1). Then, the waves propagate through the magnetic waveguides and interfere at the point of waveguide junction.  For any junction with an odd number of interfering waves, there is a transmitted wave with a non-zero amplitude. The phase of the wave passing through the junction always corresponds to the majority of the phases of the interfering waves (for example, the transmitted wave will have phase 0, if there are two or three waves with initial phase 0; the wave will have a π-phase otherwise). The transmitted wave passes the phase shifter and accumulates an additional π-phase shift (e.g. phase 0→ π, and phase π→ 0). Finally, the spin wave signal reaches the output ME cell. The output cell has two stable magnetization states. At the moment of spin wave arrival, the output cell is in the metastable state (magnetization is along the hard axis perpendicular to the two stable states).  The *phase* of the incoming spin wave defines the direction of the magnetization relaxation in the output cell [8, 10]. The process of magnetization change in the output ME cell is associated with the change of electrical polarization in the multiferroic material and can be recognized by the induced voltage across the ME cell (e.g. +10mV correspond to logic state 0, and -10mV correspond to logic 1).

The key feature of the described circuit is the combination of the digital (bi-stable) inputs/outputs and the analog core. Information processing within the analog core is associated with the manipulation of the phases of the propagating waves. The Truth Table inserted in Fig.1 shows the input/output phase correlation. The waveguide junction works as a Majority logic gate. The amplitude of the transmitted wave depends on the number of the in-phase waves, while the phase of the transmitted wave always corresponds to the majority of the phase inputs. The π-phase shifter works as an Inverter in the phase space. As a result of this combination, the three-input one-output gate in Fig.1 can operate as a NAND or a NOR gate for inputs A and B depending on the third input C (NOR if C=1, NAND if C=0). Such a gate can be a universal building block for any Boolean logic gate construction. Most importantly is that the analog core can operate on a number of frequencies at the same time by exploiting wave superposition.

The general view of the multi-frequency magnonic circuit is shown in Fig.2. The structure and the principle of operation are similar to the above described example except there are multiple ME cells on each of the input and output nodes. These cells are aimed to operate (excite and detect) spin waves on different frequencies (e.g. $f_1$, $f_2$, ... $f_n$). The frequency excited by the ME cell depends on many factors and can be adjusted by the cell size/shape/composition. In order to avoid the crosstalk among the cells operating on different frequencies, the cells are connected with the spin wave buses via the magnonic crystals [11] serving as frequency filters. Each of these crystals allows spin wave transport within a certain frequency range enabling ME cell frequency isolation. Within the spin wave buses, spin waves of different frequencies superpose, propagate, and receive a π-phase shift independently of each other. Logic 0 and 1 are encoded into the phases of the propagating spin waves on each frequency. The output ME cells are connected to the spin wave buses via the magnonic crystals in order to receive spin wave signal on the specific frequency. The Truth Table shown in Fig.1 can be applied for the each of the operating frequencies. Thus, the considered circuit can perform NAND or NOR operations on the number of bits at the same time.

III. Numerical simulations

In order to illustrate the principle of operation, we present the results of numerical simulations. The operation of the magnonic logic circuit includes several major steps: spin wave generation by the input ME cell, spin wave propagation and interference, a π-phase shift, and the output ME cell magnetization switching depending the phase of the incoming spin wave. The process of the spin wave excitation by the ME cell is modeled by using the Landau-Lifshitz equation:

$$\frac{d\vec{m}}{dt} = -\frac{\gamma}{1+\alpha^2}\vec{m}\times\left[\vec{H}_{eff} + \alpha\vec{m}\times\vec{H}_{eff}\right], \quad (1)$$

where $\vec{m} = \vec{M}/M_s$ is the unit magnetization vector, $M_s$ is the saturation magnetization, $\gamma$ is the gyro-magnetic ratio, and $\alpha$ is the phenomenological Gilbert damping coefficient. The effective magnetic field $\vec{H}_{eff}$ is the sum of the following:

$$\vec{H}_{eff} = \vec{H}_d + \vec{H}_{ex} + \vec{H}_a + \vec{H}_b, \quad (2)$$

where $\vec{H}_d$ is the magnetostatic field, $\vec{H}_{ex}$ is the exchange field, $\vec{H}_a$ is the anisotropy field ($\vec{H}_a = (2K/M_s)(\vec{m}\cdot\vec{c})\vec{c}$, $K$ is the uniaxial anisotropy constant, and $\vec{c}$ is the unit vector along the uniaxial direction), $\vec{H}_b$ is the external bias magnetic field. It is assumed that the application of the bias voltage $V$ to the ME cell results in the easy-axis rotation ($\vec{c}$ vector rotation) as follows:

$$c_x = 0, \; c_y = \cos\theta, \; c_z = c_0 \sin\theta,$$
$$\theta = \frac{\pi}{2}\left(\frac{V}{V_G}\right), \quad (3)$$

where $V_G$ is the voltage resulting in a 90 degree easy axis rotation from the Y axis towards the Z axis. The material parameters taken in simulations are typical for permalloy films $\gamma=2.0\times10^7$rad/s/Oe, $4\pi M_s$=10kG, $2K/M_s$=4Oe [12, 13]. The plot in Fig.3 shows the magnetization of the ME cell (black curve) and the amplitude of the excited spin wave (red curve) as a function of time. The plot illustrates two possible magnetization responses to the positive and negative input voltages (dome-shaped voltage pulses of 300ps). The key feature we want to illustrate is that the polarity of the applied voltage defines the direction of the ME cell's magnetization change and the initial phase (0 or π) of the excited spin wave.

The propagation, interference and phase accumulation are simulated by the analytical model described in [14]. This simple model is found to be in a good agreement with experimental data on spin wave propagation in permalloy [14]. We assume that each of the input ME cell generates a spin wave packet propagating along the X axis. The wave packet consists of a Gaussian distribution of wave vectors that is $2/\delta$ in width and centered about $k_0$. The magnetization components in the Cartesian coordinates are given as follows:

$$M_x = \frac{C\exp(-t/\tau)}{\delta^4 + \beta^2 t^2}\exp\left[\frac{-\delta^2(x-vt)^2}{4(\delta^4 + \beta^2 t^2)}\right]\sin(k_0 x - \omega t + \phi),$$

$$M_y = \frac{C\exp(-t/\tau)}{\delta^4 + \beta^2 t^2} \exp\left[\frac{-\delta^2(x-vt)^2}{4(\delta^4 + \beta^2 t^2)}\right]\cos(k_0 x - \omega t + \phi),$$
$$M_z = \sqrt{M_s^2 - M_x^2 - M_y^2}, \qquad (4)$$

where $C$ is a constant proportional to the amplitude, $\tau$ is the decay time, $\phi$ is the initial phase, $v$ and $\beta$ are the coefficients of the first and second order terms, respectively, in the Taylor expansion of the nonlinear dispersion $\omega(k)$. In this case, the propagation of the spin wave packet can be described by just one magnetization component ($M_x$ or $M_y$). Hereafter, we present the results of numerical modeling for just one magnetization component perpendicular to the spin wave propagation direction - $M_y$. In our simulations, we take $\tau$=1.0ns, $\delta$ = 1.0 μm, $v$=$10^4$m/s, and neglect the second order term $\beta$=0 for simplicity.

First, we simulate the operation of the single-frequency circuit ($k_0$=0.25 μm$^{-1}$). In Fig.4, there are several plots showing spin wave signals excited by the input ME cells (on the left), signal after the point of waveguide junction (in the center), and the spin wave signal after the phase shifter (on the right). Three input ME cells generates three spin wave packets of the shape and distribution except the initial phase - $\phi$. As an example, we took the initial phase $\phi$=0 for inputs A and C, and the opposite phase $\phi$=π for input B. The amplitudes of the spin waves are normalized to the saturation magnetization $M_s$. We intentionally restrict our consideration by the relatively small amplitudes $M_y$<<$M_s$ to consider only the linear regime. Thus, at the point of waveguide junction, the total magnetization can be found as a sum of the three superposing packets form inputs A, B, and C. The amplitude of the transmitted signal may vary depending the number of in-phase waves (two or three) while the phase of the propagating wave ($\phi$=0) corresponds to the majority of the phases (e.g. MAJ(0,π,0)=0). The wave propagating through the phase shifter signal accumulates an additional π-phase shift (the length of the inverter is taken to be 100nm).

The process of the output ME cell switching is illustrated in Fig.5. At the moment of switching, the output ME cell is polarized along the Z axis (the $V_G$ voltage applied to the output ME cell). As the spin wave packet reaches the output ME cell, the bias voltage is turned off, and the magnetization starts to relax towards the stable state along or opposite to the Y axis. The phase of the incoming wave defines the way of relaxation and the final magnetization state of the output ME cell (e.g. $\phi$=0 will lead to the positive $M_y$, and $\phi$=π will lead to the negative $M_y$). There are two curves in Fig.5 showing two possible relaxation trajectories depending on the phase of the incoming wave. We want to emphasize that the way of ME switching is determined by the phase not the amplitude of the spin wave signal. A more detailed simulations on ME switching as a function of the phase/amplitude of the incoming spin wave can be found in Ref. [10].

The results of numerical modeling shown in Figs.3-5 are aimed to illustrate the main idea of spin wave circuit – logic functionality by manipulating the *phases* of the propagating waves. There are no non-linear devices (switches) in the analog core but passive elements (waveguide junction) exploiting wave interference and phase shifters.

The latter let us extend this approach for multiple waves simultaneously propagating and interfering in one structure.

Next, we simulate the operation of the multi-frequency circuit shown in Fig.2. For simplicity, we consider three waves for each input A, B and C. We assume three ME cells on each input to generate spin wave packets centered about the three wave vectors $k_1$, $k_2$, and $k_3$, corresponding to the three operating frequencies $f_1$, $f_2$, and $f_3$, respectively. As an example, we take $k_1= 0.02\mu m^{-1}$, $k_2=0.25\mu m^{-1}$, and $k_3 =2.0\mu m^{-1}$. The phase velocity $v$, damping constant $\tau$, and the packet width $\delta$ are the same as in the previous example of the single-frequency circuit. In Fig.6, there are plots showing spin wave signals excited by the ME cells (on the left), signal after the point of waveguide junction (in the center), and the signal after the phase shifter (on the right). The spin wave signal on each input is the sum of three packets centered about three different wave vectors. Each plots on the left side of Fig.6 shows three curves corresponding to the three packets, ant the insert on the plot depicts the sum of three. As in the previous example, the information is encoded in the initial phase for each packet ($2^3$ possible phase combinations for each input, $8^3$ possible combinations for 3 inputs). The resultant magnetization at the point of waveguides junction is calculated as a sum of the nine input packets (shown in the inserts in Fig.6). The wave passing through the junction comprises three wave packets with $2^3$ possible phase combinations. It is assumed that each of the packets accumulates a π-phase shift independently of others. There are three output ME cells receive signal on one of the information carrying frequencies ($f_1$, $f_2$, $f_3$) to recognize and store one of the $2^3$ final states. This numerical illustrations are based entirely on the wave superposition. Theoretically, it is possible to build a classical wave-based core, which can be in the superposition of $2^N$ possible states, where $N$ is the number of the input ME cells. The $N$ bits can divided between $n_f$ frequencies depending on the number of inputs per logic gate $\alpha$, $N=n_f \times \alpha$. (e.g. three input bits per frequency for the universal gate shown in Fig.1).

## IV. Discussion

The principle of operation of the multi-frequency magnonic circuit is fundamentally different from the conventional transistor-based circuits and combines a number of novel ideas on information transmission and processing. Some of the circuit components have been experimentally realized (e.g. spin wave interferometer [15]) and some components (e.g. ME cells) are under study. The major critical concerns on the practical feasibility of the multi-frequency approach can be divided into the following: (i) ability to excite and recognize multiple spin waves on different frequencies, (ii) power dissipation increase due to the number of operating channels, and (iii) fault tolerance of the phase-based circuits.

ME cell is the element aimed to perform read-in and read-out operations on specific frequencies. The operation of the ME cell is based on the effect of magneto-electric coupling enabling magnetization control by applying an electric field and vice versa. For a long time, the most of interest on magneto-electric coupling has been associated with the single-phase multiferroics – a unique class of materials inherently possessing electric and magnetic polarizations (e.g. $BiFeO_3$)[16]. However, the limited number of room temperature single-phase multiferroics as well as the relatively low magnetic polarization makes difficult their practical utilization in magnonic circuits. On the other hand, a combination of piezoelectric and magnetostrictive materials (so called synthetic multiferroics) may provide the similar effect. An electric field applied across the piezoelectric produces stress, which, in turn, affects the magnetic polarization of the magnetostrictive material. As a result of the electro-mechanical-magnetic coupling the magnetization of the magnetostrictive materials can be controlled by the electric field. A number of piezo-magnetostrictive pairs have been studied during the past decade and magneto-electric coupling coefficients have been tabulated [17]. In the specific case of magnonic circuitry, it is important to produce significant change of magnetization in order to excite spin waves. Recent experimental data on PZT/Ni pair have shown the 90 degree magnetization rotation in Ni as a function of the electric field applied across the PZT layer [18]. The utilization of the synthetic multiferroics has a big advantage for the ME cell engineering as the frequency response of the cell can be controlled the thickness of the piezoelectric/magnetostrictive materials, which makes it possible to realize a frequency-selective input/output elements based on the same materials. It is important to note that the electric field required for magnetization rotation in Ni/PZT synthetic multiferroic is about 1.2MV/m [18]. The latter promises a very low, order of aJ, energy per switch achievable in nanometer scale ME cells (e.g. 24aJ for 100nm×100nm ME cell with 0.8μm PZT). Thus, the maximum power dissipation density per $1\mu m^2$ area circuit operating at 1GHz frequency can be estimated as $7.2W/cm^2$ (three input cells per one frequency). An addition of an extra operating frequency would linearly increase the power dissipation in the circuit. The upper limit for the power dissipation in magnonic circuits may be higher than the one of the silicon counterparts, as the metallic waveguides can be placed on the non-magnetic metallic base (e.g. Cu) to enhance the thermal transport.

Reliable operation of the output ME cells is another important challenge. The output cell relaxation is triggered by the seed magnetization change provided by the all incoming spin waves. The increasing number of the transmitted signals will make difficult the precise filtering of the spin wave signal. The error immunity of the read-out is directly related to the quality of the magnonic crystals [11] serving as frequency filters for the ME cells operating on different frequencies. Magnonic crystals can be fabricated as a composition of two materials with different magnetic properties or as a single material waveguide with periodically varying dimensions. Frequency band gaps have been experimentally observed in a grating-like structure comprising shallow grooves etched into the surface of an yttrium-iron-garnet film [19], in a one-dimensional arrays of permalloy nanostripes separated by the air gaps [20], and in a synthetic nanostructure composed of two different magnetic materials [21]. The obtained data show the feasibility of using magonic crystals as the frequency filters with frequency band gaps of several GHz, which can be further tuned by the bias magnetic field [21].

Spin wave transport in nanometer scale magnetic waveguides has been intensively studied during the past decade [13, 14, 22, 23]. The coherence length of spin waves in ferromagnetic materials (e.g. NiFe) exceeds tens of microns at room temperature [13, 14], which allows to utilize spin wave interference at the micrometer scale. The typical spin wave group velocity is $10^4$m/s and the relaxation time is about 0.8ns for NiFe at room temperature. These short attenuation time and relatively slow propagation are not critical for the sub-micrometer scale logic devices. It is more important to have a mechanism for a fast phase modulation. The phase of the propagating spin wave can be controlled by an external magnetic field (e.g. produced by the pinned layer) or by the waveguide thickness/length variation. However, in these cases, the time required for the π-phase change is proportional to the strength of the magnetic field or the length of the waveguide. A more promising approach for phase modulation is in the use of a domain wall of a special shape [24], which may provide a frequency-independent π-phase change within a nanometer scale length.

There are certain concerns associated with the reliability of information encoding in the phase of the propagating wave. As noticed in Ref. [25], the phase can be disrupted by scattering of the waveguide imperfections. Dispersion may be another serious problem as the velocity of the spin wave is frequency-dependent. Some of these critical comments have been clarified by the recent experimental data on three-wave interference in 20nm thick permalloy film [26]. Three spin waves were excited by microwave (3GHz) signals produced by the AC electric currents. The initial phases (0 or π) were controlled by the direction of the excitation current. All $2^3$ phase combination have been tested and reliable output detected [26]. This experiment have validated the feasibility of using spin wave interference and demonstrated robust operation at room temperature. In general, the fault tolerance of the wave-based devices is defined by the operating wavelength *λ*. A propagating wave is not affected by the structure imperfection which size is much less than the wavelength. So far, all of the experimentally demonstrated spin wave prototypes operate with the micrometer scale waves [26-28]. It is not clear if the wavelength scaling down (~100nm) will lead to significant issues.

Multi-frequency operation implies the simultaneous transmission of a number of spin waves through the ferromagnetic waveguides. The inevitable frequency mixing due to the inherent magnetic non-linearity will result in the appearance of the spurs in the frequency domain. The latter may be a serious issue limiting the bandwidth and the number of operating frequencies. However, it may be possible to keep the amplitude of the spur signals much below the main signal level. We would like to refer to the experimental data on the simultaneous spin wave excitation by two microwave signals on different frequencies reported in [29]. Indeed, there were observed output signals on the mixed frequencies (e.g. *$f_1+f_2$*, *$f_1-f_2$*, *$2f_1-f_2$*, etc.) But the amplitude of the mix frequency signals was about 20dB less than the amplitude of the main signals. A multi-frequency spin wave transport is an unexplored field, which requires a more detailed experimental study. To the best of our knowledge, there is no systematic study on the bandwidth and the frequency mixing in the ferromagnetic waveguides, which is the subject for the further investigations.

Finally, we would like to estimate the potential functional throughput enhancement due to the use of multiple frequencies and compare it with the conventional CMOS. In Fig.7(A), we present the estimates on the functional throughput in [Ops/ns cm$^2$] for the Full Adder Circuit built of scaled CMOS (blue markers) and magnonic multi-frequency circuit (red markers). The estimates for the CMOS Full Adder circuit are based on the data for the 32nm CMOS technology (area = 3.2μm$^2$, time delay= 10ps from Ref. [30]). The estimates for further generations are extrapolated by using the following empirical rule: the area per circuit scales as ×0.5 per generation, and the time delay scales as ×0.7 per generation. The estimates for the magnonic circuit are based on the model for the single frequency operating circuit presented in Ref. [10]. The area per circuit scales as $25×\lambda$, where $\lambda$ is the wavelength. The time delay $t_{delay}$ is the sum of the following: the time required to excite spin wave, the propagation time, and the time required for the output ME cell switchimng: $t_{delay} = t_{ext} + t_{prop} + t_{relax}$. The propagation time is easy to estimate by dividing the circuit length by the spin wave group velocity $t_{prop} = 3 \lambda /v$ (e.g. 100nm per 10ps). Less reliable are the estimates on the ME cell excitation and relaxation times due to the lack of experimental data. As a conservative estimate, we take $t_{ext} = t_{relax}$ =100ps, which is experimentally observed in magnetic memory devices [31]. The graph in Fig.7(A) shows significant (two orders of magnitude) functional throughput enhancement over the scaled CMOS. The overall advantage is due to the two parts (i) magnonic logic can outperform transistor-based approach even for the single frequency circuits, as the wave-based logic circuits requires a fewer number of elements [10]; (ii) an additional advantage comes from the use of multiple frequencies. The plot in Fig.7(B) shows the relative functional throughput enhancement $F_N$ as a function of the number of frequencies $N$ (independent information channels). The enhancement is estimated by the following formula:

$$F_N = F_1 + \frac{N}{(1+\frac{\Delta s}{s}N)\cdot(1+\frac{\Delta t}{t}N)}, \qquad (5)$$

where $F_1$ is the functional throughput for a single frequency, $\Delta s/s$ is the relative area increase associated with the addition of one extra frequency, the $\Delta t/t$ is the relative time delay increase associated with the addition of one extra frequency. In numerical estimates, we assumed $\Delta s/s$ =5%, and $\Delta t/t$=1%. As one can see from the Fig.5(B), the relative enhancement of using multiple frequencies estimated by Eq. (2) has a maximum and starts to decrease beyond the optimum channel number. The decrease at large number of channels is due to the additional area and time delay introduced by each new input/output port. The optimum number of the operational channels may vary for different logic circuits. The lower is the area/time delay per an additional input/output port the more prominent is the advantage of the multi-frequency circuits.

**V. Conclusions**

We have described a concept of multi-frequency magnonic logic circuits taking the advantage of wave superposition for parallel data processing. The operation is illustrated by numerical modeling. According to the presented estimates, mutli-frequency magnonic logic circuits may provide an orders of magnitude functional throughput enhancement over the scaled CMOS. There is a number of questions regarding the operation of the circuit components and overall system stability, which require further study. In summary, we want to emphasize the key point of this work. Wave-based magnonic logic devices offer a fundamental advantage over the CMOS technology. The ability to use multiple frequencies as independent information channels opens a new dimension for functional throughput enhancement and may provide a route to a long-term development.

## Acknowledgments

The work was supported by the DARPA program on Non-volatile Logic (program manager Dr. Devanand K. Shenoy) and by the Nanoelectronics Research Initiative (NRI) (Dr. Jeffrey J. Welser, NRI Director) via the Western Institute of Nanoelectronics (WIN , director Dr. Kang L Wang).

## References

[1]    G. E. Moore, Electronics 38 (1965) 114.
[2]    V. A. Sverdlov, T. J. Walls, and K. K. Likharev, IEEE Transactions on Electron Devices 50 (2003) 1926.
[3]    K. Bernstein, R. K. Cavin, W. Porod, A. Seabaugh, and J. Welser, Proceedings of the IEEE 98 (2010) 2169.
[4]    P. R. Bandaru, C. Daraio, S. Jin, and A. M. Rao, Nat Mater 4 (2005) 663.
[5]    A. K. Geim, Science 324 (2009) 1530.
[6]    Q. Zhang, T. Fang, H. Xing, A. Seabaugh, and D. Jena, IEEE Electron Device Lett. 29 ( 2008) 1344.
[7]    A. Khitun and K. Wang, Superlattices & Microstructures 38 (2005) 184.
[8]    A. Khitun, B. M., and W. K.L., IEEE Transactions on Magnetics 44 (2008) 2141.
[9]    A. Khitun, M. Bao, and K. L. Wang, Superlattices & Microstructures 47 (2010) 464.
[10]   A. Khitun and K. L. Wang, http://arxiv.org/abs/1012.4768 (2010)
[11]   M. Krawczyk and H. Puszkarski, Journal of Applied Physics 100 (2006) 073905.
[12]   W. K. Hiebert, A. Stankiewicz, and M. R. Freeman, Physical Review Letters 79 (1997) 1134.
[13]   T. J. Silva, C. S. Lee, T. M. Crawford, and C. T. Rogers, Journal of Applied Physics 85 (1999) 7849.
[14]   M. Covington, T. M. Crawford, and G. J. Parker, Physical Review Letters 89 (2002) 237202.
[15]   Y. Wu, M. Bao, A. Khitun, J.-Y. Kim, A. Hong, and K.L. Wang, Journal of Nanoelectronics and Optoelectronics 4 (2009) 394.
[16]   S.-W. Cheong and M. Mostovoy, Nat Mater 6 (2007) 13.

[17] W. Eerenstein, N. D. Mathur, and J. F. Scott, Nature 442 (2006) 759.
[18] T. K. Chung, S. Keller, and G. P. Carman, Applied Physics Letters 94 (2009)
[19] A. V. Chumak, A. A. Serga, B. Hillebrands, and M. P. Kostylev, Appl. Phys. Lett. 93 (2008) 022508.
[20] M. P. Kostylev, G. Gubbiotti, J.-G. Hu, G. Garlotti, T. Ono, and R. L. Stamps, Phys. Rev. B 76 (2007) 054422.
[21] Z. K. Wang, V. L. Zhang, H. S. Lim, S. C. Ng, M. H. Kuok, S. Jain, and A. O. Adeyeye, Appl. Phys. Lett. 94 (2009) 083112.
[22] M. Bailleul, D. Olligs, C. Fermon, and S. Demokritov, Europhysics Letters 56 (2001) 741.
[23] V. E. Demidov, J. Jersch, K. Rott, P. Krzysteczko, G. Reiss, and S. O. Demokritov, Physical Review Letters 102 (2009) 177207.
[24] I. V. Ovchinnikov and K. L. Wang, PHYSICAL REVIEW B 82 (2010) 024410.
[25] S. Bandyopadhyay and M. Cahay, Nanotechnology 20 (2009) 412001.
[26] P. Shabadi, A. Khitun, P. Narayanan, M. Bao, I. Koren, K. L. Wang, and C. A. Moritz, Proceedings of the Nanoscale Architectures (NANOARCH), 2010 IEEE/ACM International Symposium (2010) 11.
[27] M. P. Kostylev, A. A. Serga, T. Schneider, B. Leven, and B. Hillebrands, Applied Physics Letters 87 (2005) 153501.
[28] Y. Wu, M. Bao, A. Khitun, J.-Y. Kim, A. Hong, and K. L. Wang, J. Nanoelectron. Optoelectron 4 (2009) 394.
[29] M. Bao, A. Khitun, Y. Wu, J.-Y. Lee, K. L. Wang, and A. P. Jacob, Applied Physics Letters 93 (2008)
[30] A. Chen, private communication (2010)
[31] G. E. Rowlands, T. Rahman, J. A. Katine, J. Langer, A. Lyle, H. Zhao2, J. G. Alzate, A. A. Kovalev, Y. Tserkovnyak, Z. M. Zeng, H. W. Jiang, K. Galatsis, Y. M. Huai, P. K. Amiri, K. L. Wang, I. N. Krivorotov, and J.-P. Wang, Appl. Phys. Lett. 98 (2011)

**Figure captions**

Fig.1 Schematic view of the single-frequency operating magnonic logic circuit. There are three inputs (A,B,C) and the output D. The inputs and the output are the ME cells connected via the ferromagnetic waveguides –spin wave buses. The input cells generate spin waves of the same amplitude with initial phase 0 or $\pi$, corresponding to logic 0 and 1, respectively. The phase is controlled by the polarity of the input voltage pulse (e.g. ±10mv). The waves propagate through the waveguides and interfere at the point of junction. The phase of the wave passed the junction corresponds to the majority of the interfering waves. The phase of the transmitted wave is inverted (e.g. passing the domain wall). The Table illustrates the data processing in the phase space. The phase of the transmitted wave defines the final magnetization of the output ME cell D. The circuit can operate as NAND or NOR gate for inputs A and B depending the third input C (NOR if C=1, NAND if C=0).

Fig.2 Schematic view of the multi-frequency magnonic circuit. There are multiple ME cells on each of the input and output node aimed to excite and detect spin waves on the specific frequency (e.g. $f_1, f_2, ... f_n$). The cells are connected to the spin wave buses via the magnonic crystals serving as the frequency filters. Within the spin wave buses, spin waves of different frequencies superpose, propagate, and receive a $\pi$-phase shift independently of each other. Logic 0 and 1 are encoded into the phases of the propagating spin waves on each frequency. The output ME cells recognize the result of computation (the phase of the transmitted wave) on one of the operating frequency.

Fig.3 (A) Spin wave excitation by the ME cell. The ME cell is a multiferroic structure enabling magnetization control by the electric field. The applied voltage rotates the direction of the anisotropy field from the Y axis toward the Z axis. (B) Results of numerical modeling showing the magnetization change of the ME cell (black curve) and the excited spin wave (red curve). The direction of magnetization rotation is defined by the polarity of the applied electric field.

Fig.4 Results of numerical simulations illustrating the operation of the single-frequency logic circuit shown in Fig.1. The plots on the right show the $M_y$ component of the spin wave signal generated by the input ME cell. The plots in the center and on the left show the signal after the point of junction and the $\pi$-phase shifter, respectively. The spin wave signals are approximated by the wave packets with Gaussian distribution (width $\delta=1\mu m$ and centered about $k_0=0.25$ $\mu m^{-1}$.

Fig.5 Output ME cell switching as a function of the phase of the incoming spin wave The ME cell is polarized along the Z axis (the $V_G$ voltage applied to the output ME cell) prior to the switching. The bias voltage is turned off at the moment of the spin wave arrival. The magnetization starts to relax towards the stable state along or opposite to the Y axis. The relaxation trajectory is defined by the phase of the incoming wave. The blue curve and the red curves show the two possible trajectories corresponding to 0 and $\pi$ phase of the incoming wave, respectively.

Fig.6 Results of numerical modeling illustrating the operation of the triple-frequencies logic circuit shown in Fig.2. The operating wave vectors are $k_1= 0.02\mu m^{-1}$, $k_2=0.25\mu m^{-1}$, and $k_3 =2.0\mu m^{-1}$. The plots on the right show the $M_y$ component of the spin wave signal generated by the input ME cell. The plots in the center and on the left show the signal after the point of junction and the $\pi$-phase shifter, respectively. The inserts show the total magnetization as a sum of all superposed packets.

Fig.7 (A) Numerical estimates on the functional throughput for the Full Adder Circuit built of CMOS (blue markers) and magnonic multi-frequency circuit (red markers). The estimates for the CMOS circuit are based on the 32nm CMOS technology. The estimates for smaller feature size are extrapolated: the area per circuit scales as ×0.5 per generation, and the time delay scales as ×0.7 per generation. The estimates for the magnonic circuit are based on the model for the single frequency operating circuit reported in Ref. [10] and Eq.(2). (B) Numerical estimates on the functional throughput as a function of the number of frequencies (independent information channels) by Eq.(2).

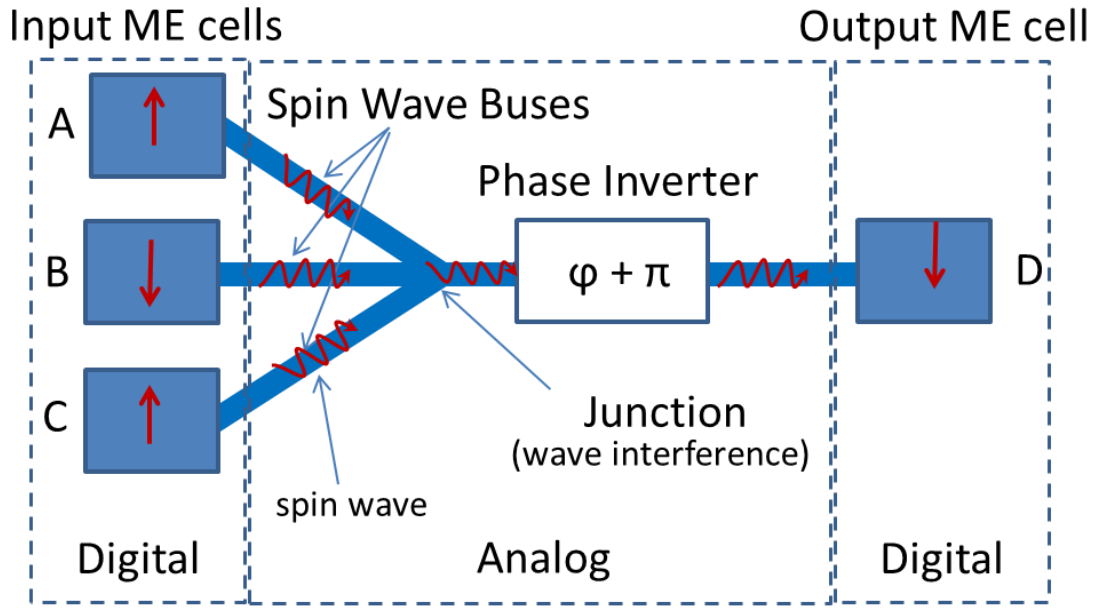

| voltage | phase |
|---|---|
| + 10mV | 0 |
| - 10mV | π |

| A | B | C | Junction | Inverter |
|---|---|---|---|---|
| 0 | 0 | 0 | 0 | π |
| 0 | 0 | π | 0 | π |
| 0 | π | 0 | 0 | π |
| 0 | π | π | π | 0 |
| π | 0 | 0 | 0 | π |
| π | 0 | π | π | 0 |
| π | π | 0 | π | 0 |
| π | π | π | π | 0 |

| phase | voltage |
|---|---|
| 0 | + 10mV |
| π | - 10mV |

**Fig.1**

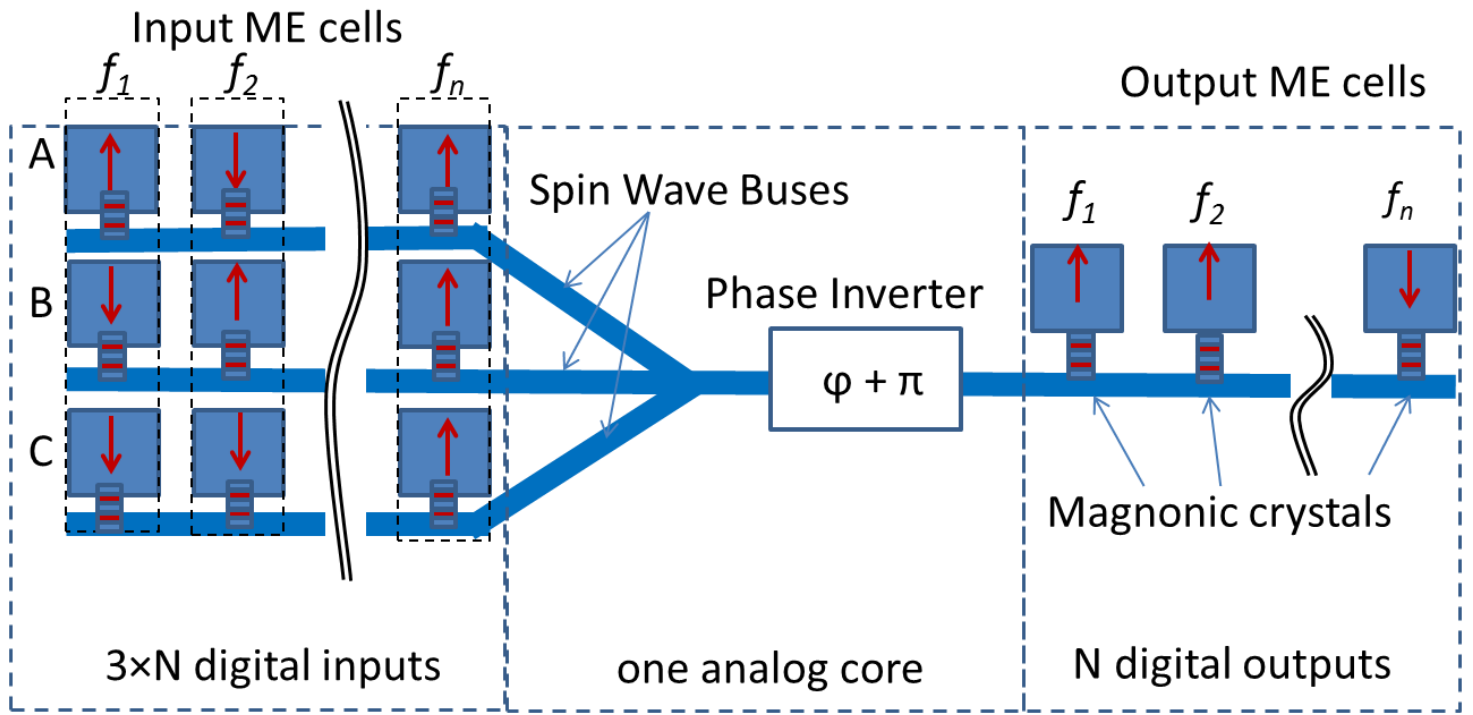

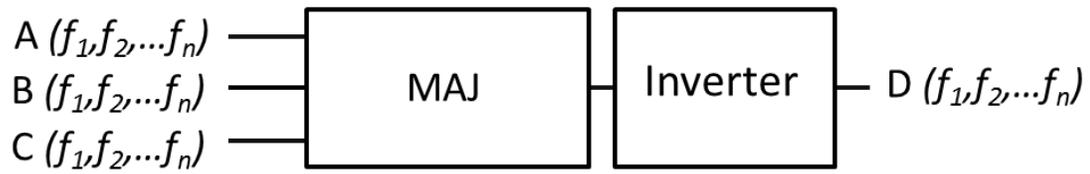

**Fig.2**

A)

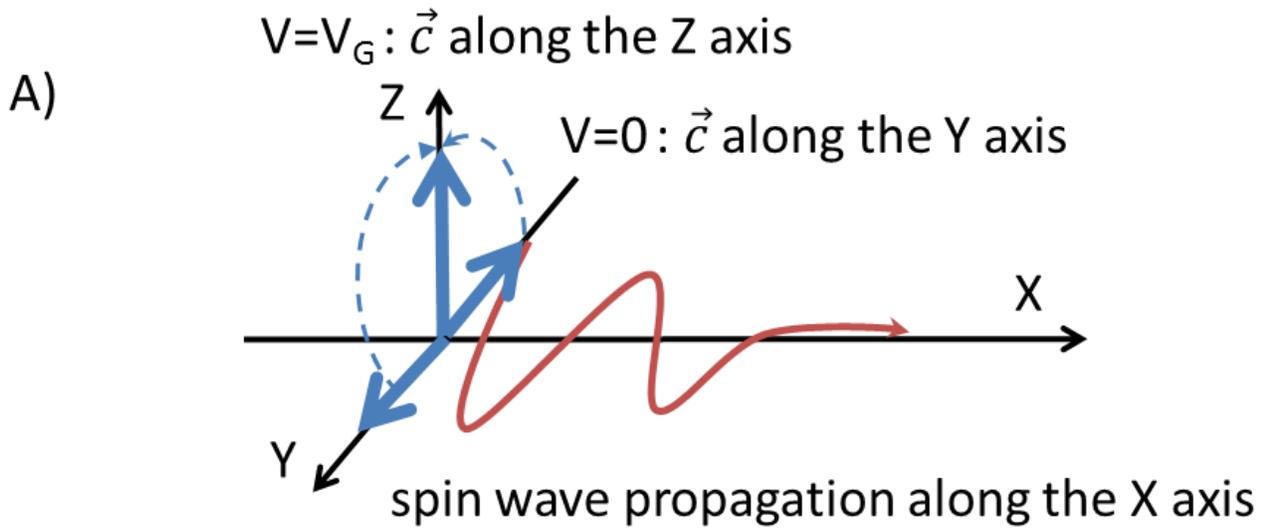

B)

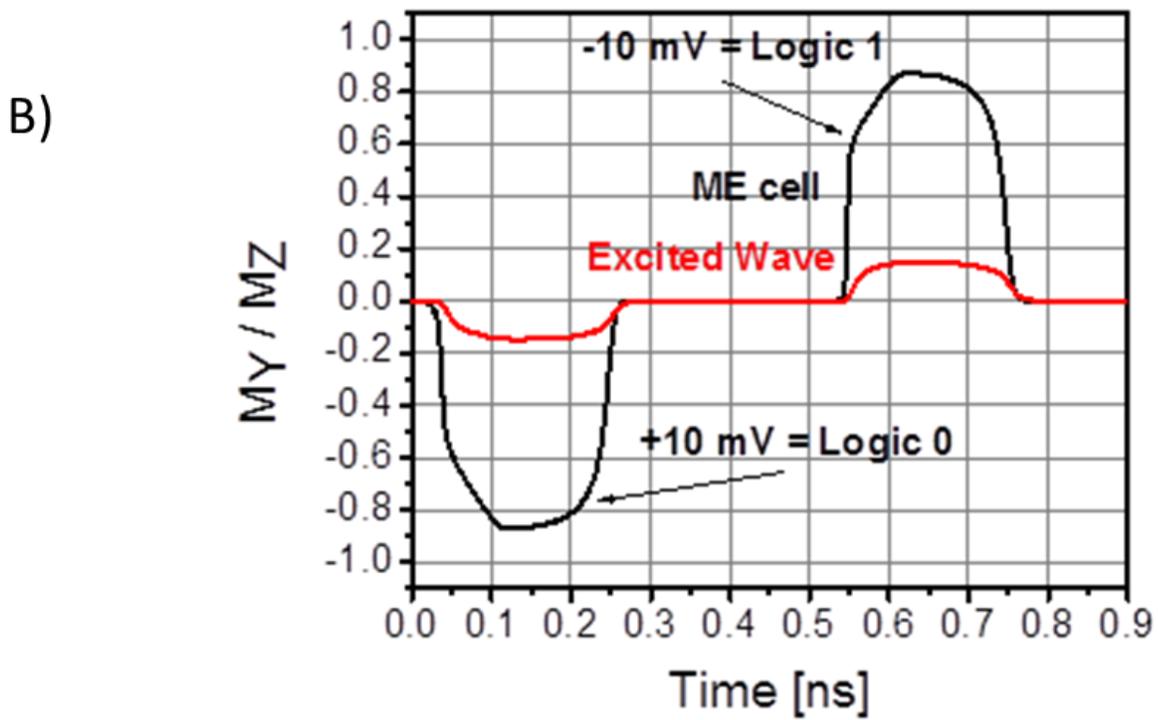

**Fig.3**

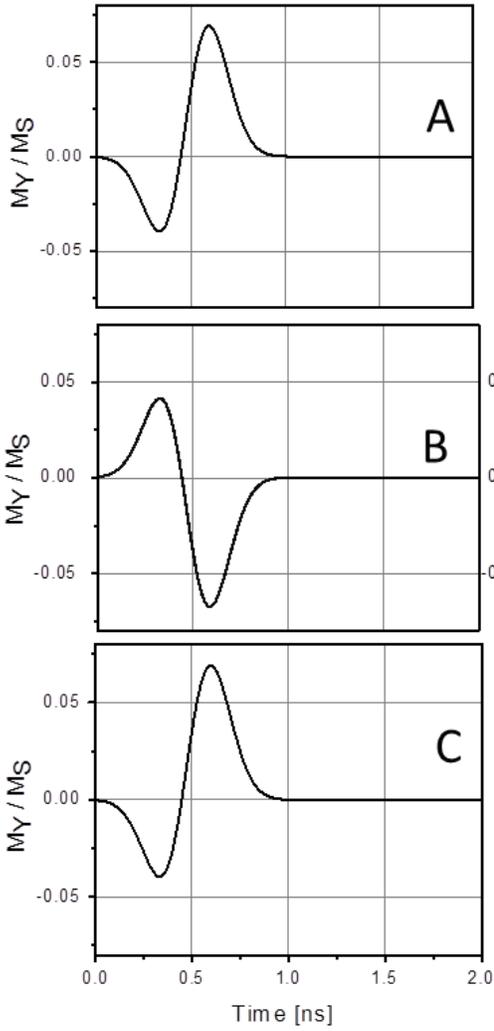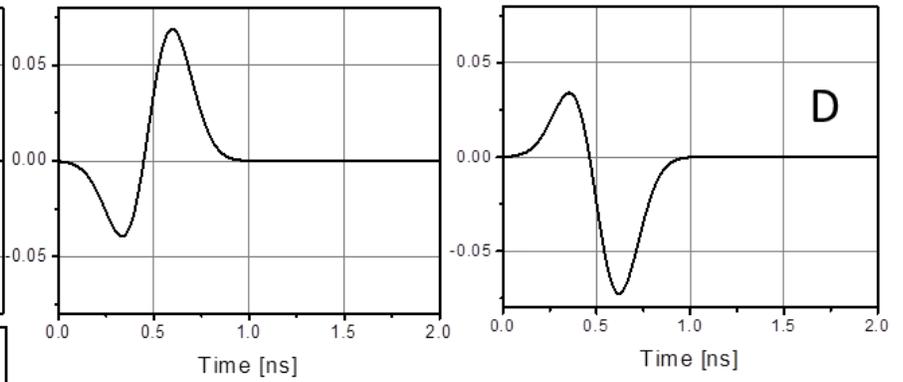

**Fig.4**

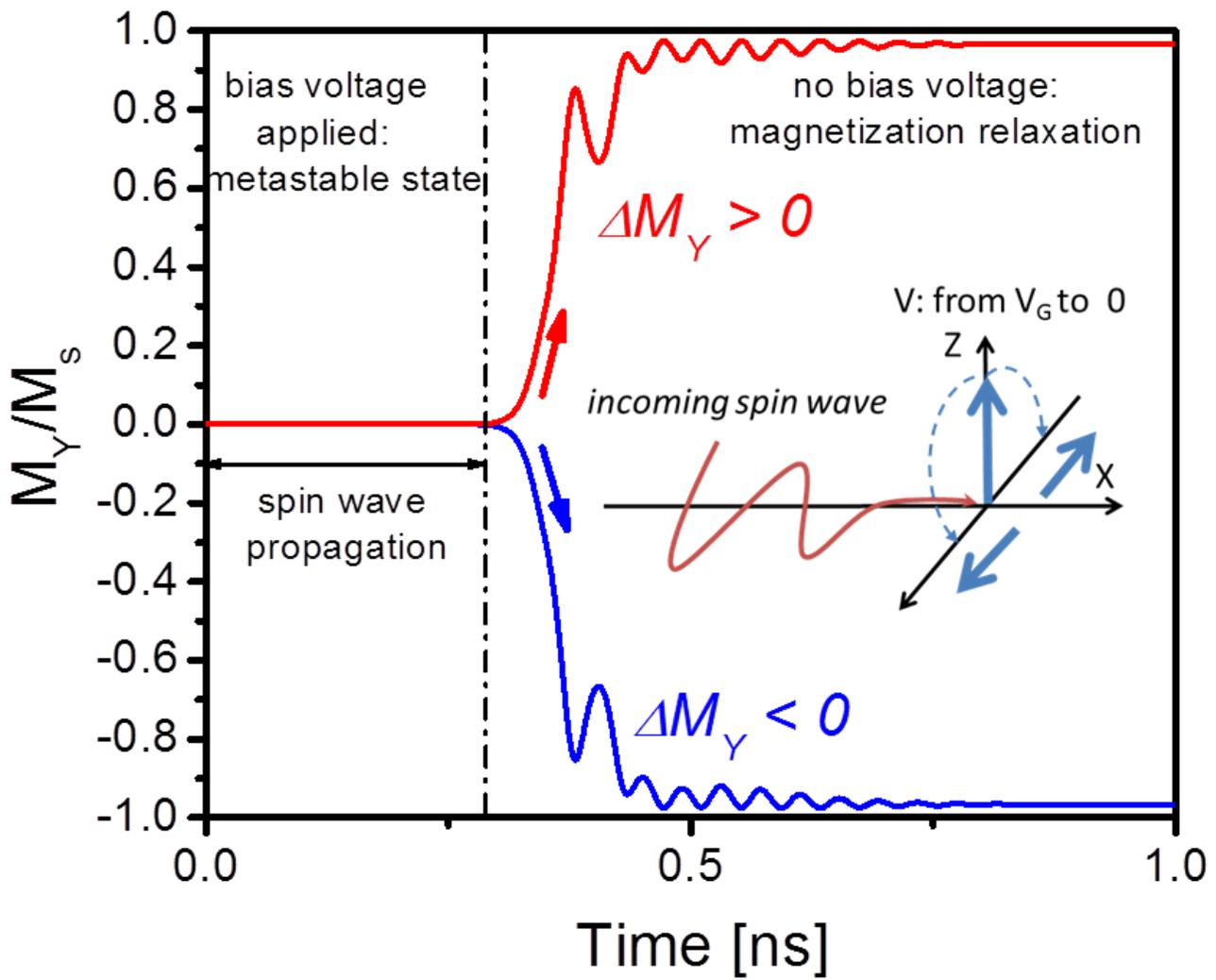

**Fig.5**

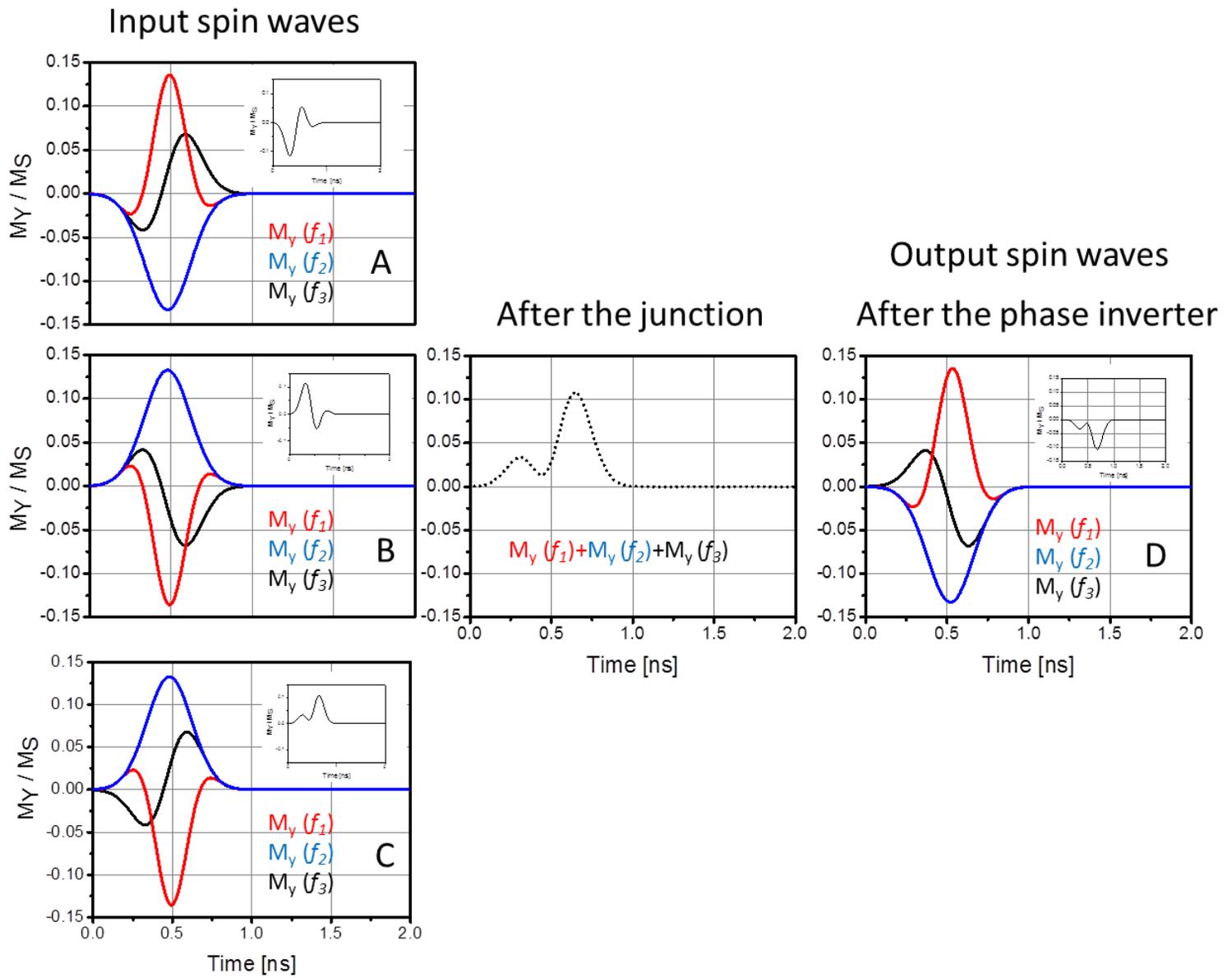

**Fig.6**

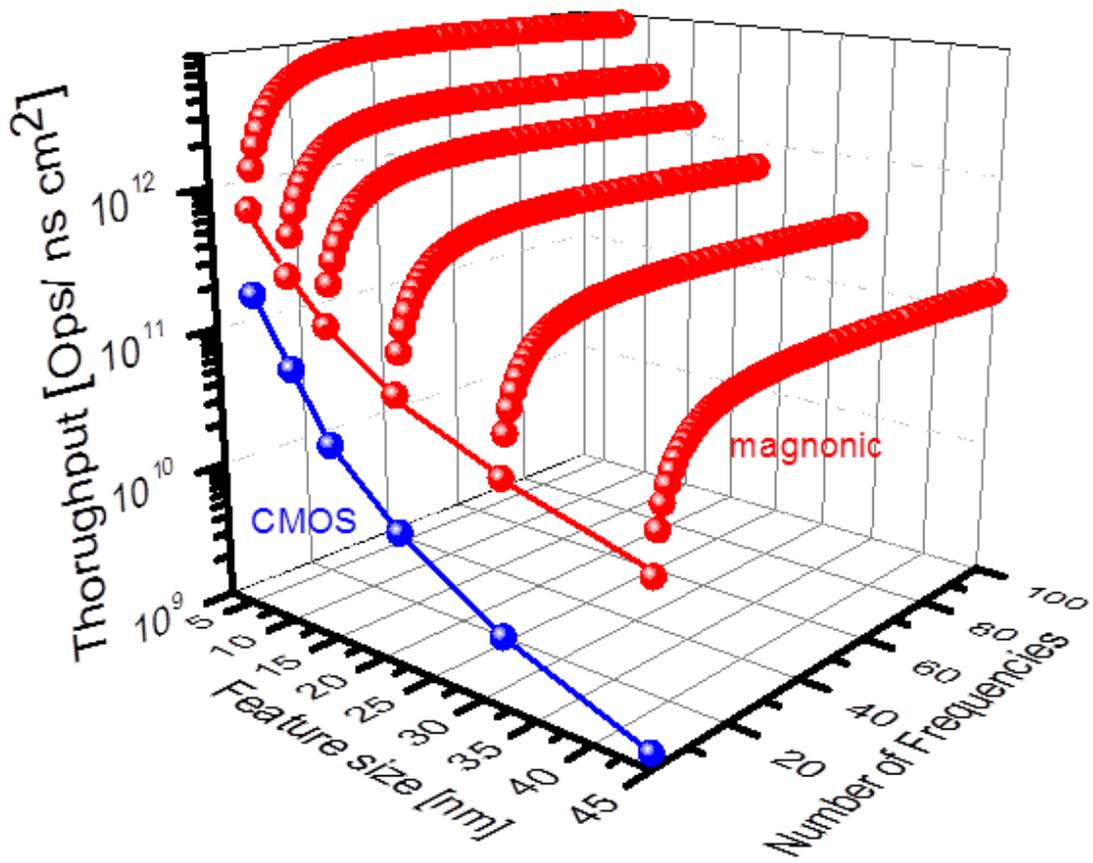

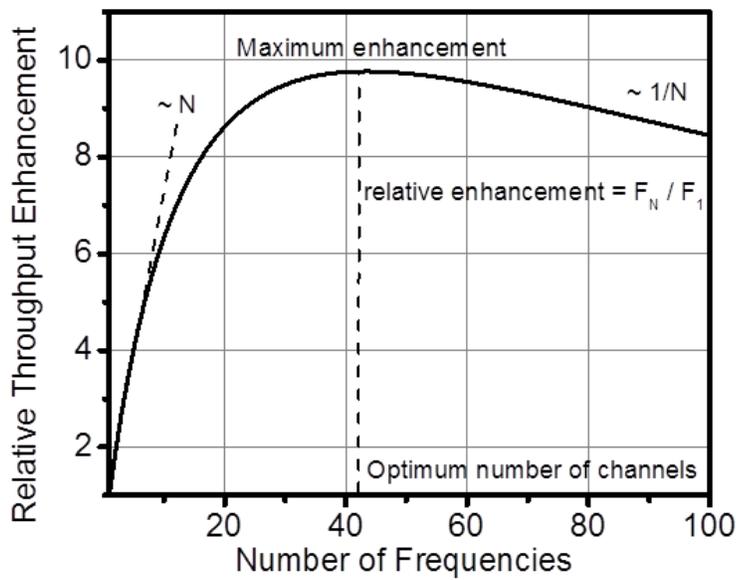

**Fig.7**